\documentclass[journal,transmag]{IEEEtran}
\usepackage{color,graphicx}

\begin{document}

\title{Static Magnetic Proximity Effect in Pt Layers on Sputter-Deposited NiFe$_2$O$_4$ and on Fe of Various Thicknesses Investigated by XRMR}

\author{\IEEEauthorblockN{Timo Kuschel\IEEEauthorrefmark{1}, Christoph Klewe\IEEEauthorrefmark{1}, Panagiota Bougiatioti\IEEEauthorrefmark{1}, Olga Kuschel\IEEEauthorrefmark{2,3}, Joachim Wollschl\"ager\IEEEauthorrefmark{2,3},\\Laurence Bouchenoire\IEEEauthorrefmark{4,5}, Simon D. Brown\IEEEauthorrefmark{4,5}, Jan-Michael Schmalhorst\IEEEauthorrefmark{1}, Daniel Meier\IEEEauthorrefmark{1}, and G\"unter Reiss\IEEEauthorrefmark{1}}

\IEEEauthorblockA{\IEEEauthorrefmark{1}Center for Spinelectronic Materials and Devices, Department of Physics, Bielefeld University, 33615 Bielefeld, Germany}
\IEEEauthorblockA{\IEEEauthorrefmark{2}Fachbereich Physik, Universit\"at Osnabr\"uck, 49069 Osnabr\"uck, Germany}
\IEEEauthorblockA{\IEEEauthorrefmark{3}Center of Physics and Chemistry of New Materials, Universit\"at Osnabr\"uck, 49069 Osnabr\"uck, Germany}
\IEEEauthorblockA{\IEEEauthorrefmark{4}XMaS, European Synchrotron Radiation Facility, Grenoble, 38043, France}
\IEEEauthorblockA{\IEEEauthorrefmark{5}Department of Physics, University of Liverpool, Liverpool, L69 7ZE, United Kingdom}
\thanks{Manuscript received November 6, 2015. 
Corresponding author: T. Kuschel (email: tkuschel@physik.uni-bielefeld.de.)}}

\markboth{IEEE TRANSACTIONS ON MAGNETICS}
{Shell \MakeLowercase{\textit{et al.}}: Bare Demo of IEEEtran.cls for IEEE Transactions on Magnetics Journals}
\IEEEtitleabstractindextext{%
\begin{abstract}
The longitudinal spin Seebeck effect is detected in sputter-deposited NiFe$_{2}$O$_{4}$ films using Pt as a spin detector and compared to previously investigated NiFe$_{2}$O$_{4}$ films prepared by chemical vapor deposition. Anomalous Nernst effects induced by the magnetic proximity effect in Pt can be excluded for the sputter-deposited NiFe$_{2}$O$_{4}$ films down to a certain limit, since x-ray resonant magnetic reflectivity measurements show no magnetic response down to a limit of $0.04\,\mu_{B}$ per Pt atom comparable to the case of the chemically deposited NiFe$_{2}$O$_{4}$ films. These differently prepared films have various thicknesses. Therefore, we further studied Pt/Fe reference samples with various Fe thicknesses and could confirm that the magnetic proximity effect is only induced by the interface properties of the magnetic material.
\end{abstract}

}

\maketitle



\section{Introduction}

\IEEEPARstart{T}{he spin Seebeck effect} \cite{Uchida:2010a} has become an important tool in spintronics \cite{Hoffmann:2015} and spin caloritronics \cite{Bauer:2012} to generate thermally driven spin currents. However, the existence of the transverse spin Seebeck effect using in-plane thermal gradients could neither be confirmed for magnetic metals (\cite{Huang:2011,Avery:2012,Schmid:2013,Meier:2013b,Shestakov:2015}), semiconductors \cite{Soldatov:2014}, nor insulators \cite{Meier:2015}. All measured response could be related to the anomalous Nernst effect (ANE) \cite{Huang:2011} or to the longitudinal spin Seebeck effect \cite{Meier:2015}, if there are unintended out-of-plane temperature gradients involved (\cite{Schmid:2013,Meier:2013b}). Recently, an additional unintended contribution of the anisotropic magnetoresistance based on inhomogeneous magnetic fields has been added \cite{Shestakov:2015} to the list of uncertainties in transverse spin Seebeck effect experiments.

The longitudinal spin Seebeck effect (LSSE) is more reliable and could be confirmed by a large number of groups (\cite{Meier:2013a,Ramos:2013,Vlietstra:2014,Siegel:2014,Gepraegs:2014}). Here, the thermal gradient is typically aligned out-of-plane. So far, the LSSE was investigated in several iron garnets (\cite{Siegel:2014,Gepraegs:2014,Uchida:2013}) and ferrites (\cite{Meier:2013a,Ramos:2013,Uchida:2010b,Li:2014,Niizeki:2015,Guo:2015}). In most cases the thermally induced spin current is detected by the inverse spin Hall effect (ISHE) \cite{Saitoh:2006} in an adjacent material such as Pt. Since Pt is close to the Stoner criterion, it can easily be spin polarized at the interface which could lead to a magnetic proximity effect (MPE) and, therefore, to an MPE induced ANE in LSSE experiments \cite{Huang:2012}.

The MPE in Pt adjacent to magnetic metals has been investigated intensely by x-ray magnetic circular dichroism (XMCD), while such studies for Pt on magnetic insulators are still rare. So far, unclear results are reported for Pt on yttrium iron garnet (\cite{Gepraegs:2012,Lu:2013}), while the MPE for Pt on CoFe$_2$O$_4$ could recently be excluded using XMCD \cite{Valvidares:2015}. 

In case of NiFe$_2$O$_4$ (NFO) deposited by chemical vapor deposition (CVD) we excluded the MPE in Pt using x-ray resonant magnetic reflectivity (XRMR) \cite{Kuschel:2015}. We further demonstrated that XRMR gives a magnetic response independent of the Pt thickness by investigating a Pt($x$)/Fe series with $x$ from $1.8\,\textrm{nm}$ to $20\,\textrm{nm}$ \cite{Kuschel:2015}. This interface-sensitivity based on the interference of reflected light at the interfaces of the bilayers is more advantageous compared to standard XMCD fluorescence measurements. This is, because the magnetic circular dichroism in reflection generates slightly different x-ray reflectivity (XRR) for opposite magnetization directions ($\pm$) due to a change of the optical constants $\pm\Delta\delta$ and $\pm\Delta\beta$ of the spin polarized material with the refractive index $n=1-\delta+i\beta$ ($\delta$:~dispersion, $\beta$:~absorption).

In this paper, we investigate the MPE in Pt on sputter-deposited NFO \cite{Klewe:2014} and use the results for the interpretation of LSSE measurements in this system. The sputter-deposited NFO films are typically thinner compared to the CVD prepared samples and have larger coercive fields. Therefore, we additionally study the thickness dependence of the MPE in Pt/Fe($x$) reference samples with $x$ from $1.1\,\textrm{nm}$ to $18.2\,\textrm{nm}$. 

\section{Experimental background and data processing}

Fe was deposited on MgAl$_2$O$_4$(001) (MAO) substrates by dc magnetron sputtering in Ar$^{+}$ atmosphere in the range of $2\times 10^{-3}\,\textrm{mbar}$. We prepared the NFO films on MAO by reactive co-sputter deposition in a pure oxygen atmosphere of the same pressure \cite{Klewe:2014} prior the in-situ sputter deposition of Pt. The magnetic moment of the sputter-deposited NFO films determined by alternating gradient magnetometer is $244\,\textrm{emu/ccm}$ in the same range as for our other sputter-deposited NFO films using the same deposition parameters \cite{Klewe:2014}. This value also agrees with the magnetic moment of the CVD prepared NFO samples as deduced from the vibrating sample magnetometer measurements presented in Ref. \cite{Kuschel:2015} (using memu instead of $\mu$emu in Fig. 1(d) of Ref. \cite{Kuschel:2015}, which was falsly labeled).

XRMR was measured at room temperature at the XMaS beamline BM28 \cite{Brown:2001} at ESRF (Grenoble, France) in $\theta-2\theta$ scattering geometry using circularly polarized x-rays with the off-resonant ($11465\,\textrm{eV}$) and resonant ($11565\,\textrm{eV}$) photon energy regarding the Pt $L_3$ absorption edge. For each angle of incidence $\theta$ an external magnetic field of $\pm200\textrm{mT}$ was applied in the scattering plane parallel to the sample surface, while the reflected intensity $I_{\pm}$ was detected. The degree of circular polarization of the x-rays was $(88\pm1)\%$ as deduced from a model to describe the performance of phase-plates \cite{Bouchenoire:2003}.

The dependence of the non-magnetic XRR intensity $I$ (zero magnetic field) and the magnetic XRMR asymmetry ratio $\Delta I=\frac{I_+-I_-}{I_++I_-}$ on the scattering vector $q=\frac{4\,\pi}{\lambda}\,\sin\theta$ ($\lambda$: wavelength) was simulated with ReMagX \cite{Macke:2014}. The off-resonant XRR curves are fitted using literature values for the optical constants. Thus, the obtained structural parameters such as thicknesses and roughnesses are used to fit the resonant XRR curves and determine the resonant optical parameters. As a last step, the resonant XRMR asymmetry ratios are fitted using the previouly obtained parameters, while varying magnetooptic profiles for the change of optical constants $\Delta \beta$ and $\Delta \delta$ \cite{Klewe:2015}. In order to obtain the magnetic moment of the Pt, we compared the change of $\Delta \beta$ and $\Delta \delta$ to theoretical calculations which have been done before \cite{Kuschel:2015}.

\section{Results and discussion}

\begin{figure}[!b]
\centering
\includegraphics[width=3in]{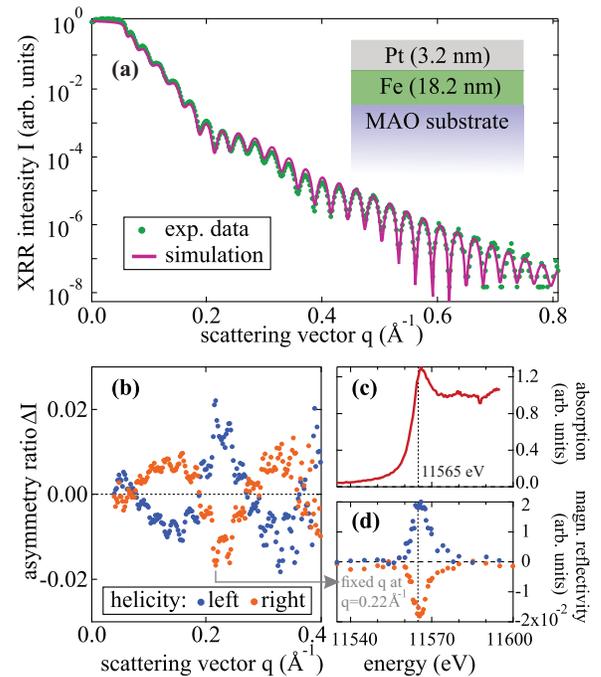}
\caption{(a)~Off-resonant ($11465\,\textrm{eV}$) non-magnetic XRR intensity $I$ and fit for Pt/Fe/MAO. (b)~Resonant ($11565\,\textrm{eV}$) asymmetry ratio $\Delta I(q)$ for both circular polarizations of the x-rays. (c)~Fluorescence spectrum at the Pt $L_3$ edge normalized to the edge jump. (d)~Magnetic reflectivity for variation of the photon energy for a fixed value of $q=0.22\,\textrm{\AA}^{-1}$.}
\label{fig:Figure1}
\end{figure}
In Fig.~\ref{fig:Figure1}(a) the off-resonant non-magnetic XRR intensity $I(q)$ of Pt/Fe/MAO shows Kiessig oscillations for Pt($3.2\,\textrm{nm}$) with roughness of $0.4\,\textrm{nm}$ and Fe($18.2\,\textrm{nm}$) with roughness of $0.5\,\textrm{nm}$. The roughness of the substrate is $0.2\,\textrm{nm}$. The resonant XRMR asymmetry ratio $\Delta I(q)$ in Fig.~\ref{fig:Figure1}(b) detected at $11565\,\textrm{eV}$ is about $2\%$ and changes sign when the helicity is reversed, which confirms the magnetic origin of the effect. The chosen energy is slightly below the maximum of the absorption edge (cf. Fig.~\ref{fig:Figure1}(c)), since here the magnetic dichroism of the spin polarized Pt is maximal \cite{Kuschel:2015}. Compared to Ref. \cite{Kuschel:2015} the energy position of the absorption maximum is slightly shifted due to small differences in the energy calibration. The whiteline intensity (ratio of absorption maximum and edge jump) in Fig.~\ref{fig:Figure1}(c) is 1.32, which indicates a mainly metallic state for Pt \cite{Kuschel:2015}. Furthermore, we varied the energy for a fixed scattering vector $q=0.22\,\textrm{\AA}^{-1}$ at a maximum position of the asymmetry ratio (cf. Fig.~\ref{fig:Figure1}(b)) as presented in Fig.~\ref{fig:Figure1}(d) to illustrate the XRMR energy dependence. The effect is maximal near the absorption maximum and vanishes at energies of more than 20 eV below and above as also confirmed by a series of XRMR scans for different energies \cite{Klewe:2015}.
\begin{figure}[!b]
\centering
\includegraphics[width=3in]{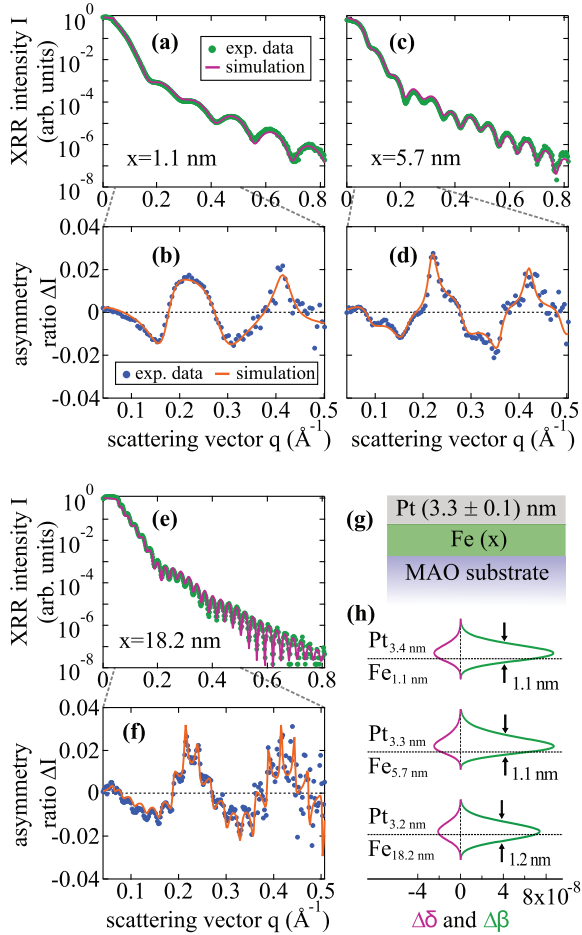}
\caption{Resonant non-magnetic XRR ($11565\,\textrm{eV}$) intensity $I$ and corresponding asymmetry ratios $\Delta I$ for Pt($3.3\pm0.1\,\textrm{nm}$)/Fe($x$) with (a),(b)~$x=1.1\,\textrm{nm}$ (average of 2 curves), (c),(d)~$x=5.7\,\textrm{nm}$ (average of 2 curves) and (e),(f)~$x=18.2\,\textrm{nm}$ (average of 4 curves). For the XRR fits the model in (g) and for the asymmetry fits the magnetooptic profiles in (h) are used.}
\label{fig:Figure2}
\end{figure}

In total, we investigated Pt($3.3\pm0.1)\,\textrm{nm}$/Fe($x$) for $x=1.1\,\textrm{nm}$, $5.7\,\textrm{nm}$ and $18.2\,\textrm{nm}$. The non-magnetic resonant XRR results are presented in Fig.~\ref{fig:Figure2}(a), (c) and (e) with similar roughnesses as mentioned before. The whiteline intensities of the absorption edges of the $1.1\,\textrm{nm}$ and the $5.7\,\textrm{nm}$ sample are even lower (1.31 and 1.27) compared to the $18.2\,\textrm{nm}$ sample. The resonant XRMR asymmetry ratios in Fig.~\ref{fig:Figure2}(b), (d) and (f) are all in the range of about $2\%$, independent of the Fe thickness. This result confirms that the Pt spin polarization is only affected by the interface properties. The XRR model in Fig.~\ref{fig:Figure2}(g) and the magnetooptic profiles in Fig.~\ref{fig:Figure2}(h) were used as input parameters in the fits. There is excellent agreement between the models and both XRR and XRMR data. The maximum in $\Delta \beta$ is slightly smaller for the thickest Fe film but all results are still comparable to previously found asymmetry ratios (\cite{Kuschel:2015,Klewe:2015}), if the degree of circular polarization is taken into account (here: $(88\pm1)\%$, Ref. (\cite{Kuschel:2015,Klewe:2015}): $(99\pm1)\%$). Since the energy for the XRMR measurements was chosen slightly below the absorption maximum, we used a fixed ratio of $\Delta\beta/\Delta\delta=-3.5$ for the fitting (cf. Ref. \cite{Kuschel:2015}). Using the calibration of Ref. \cite{Kuschel:2015}, we obtain a maximum Pt magnetic moment of $(0.5\pm0.1)\,\mu_{B}$ for all samples.
\begin{figure}[!t]
\centering
\includegraphics[width=3in]{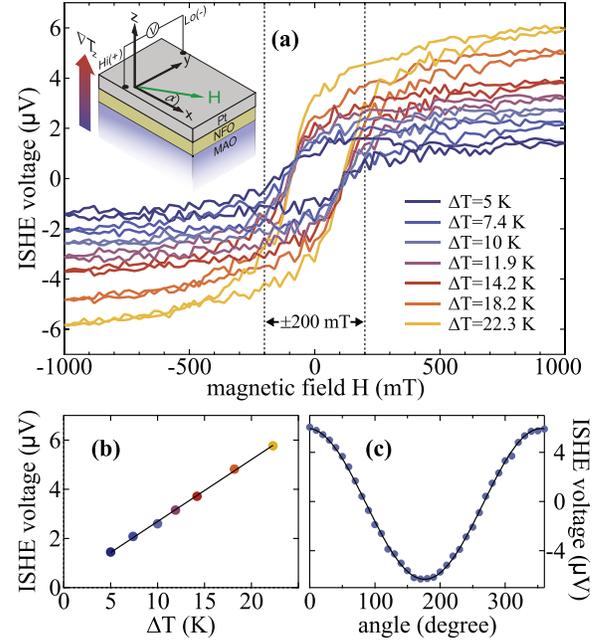}
\caption{LSSE measurements for Pt/NFO/MAO. (a)~ISHE voltage for various temperature differences $\Delta T$. Inset: Measurement geometry. (b)~Saturation value of the ISHE voltage depending on $\Delta T$. (c)~Angular dependence of the ISHE voltage in magnetic saturation ($H=1000\,\textrm{mT}$, $\Delta T=23\,\textrm{K}$).}
\label{fig:Figure3}
\end{figure}

The LSSE curves of the sputter-deposited NFO sample are presented in Fig.~\ref{fig:Figure3}(a) for various temperature gradients. These magnetic field loops have larger coercive fields \cite{Klewe:2014} compared to CVD prepared NFO \cite{Kuschel:2015}. However, we still reach $70\%$ of magnetic saturation in the XRMR experiments using the magnetic field of $\pm200\,\textrm{mT}$ which is sufficient to observe dichroic effects. The linear dependence of the LSSE on the temperature difference (cf. Fig.~\ref{fig:Figure3}(b)) and the typical cosine angular dependence in magnetic saturation (cf. Fig.~\ref{fig:Figure3}(c)) confirm the LSSE being of the same order as for CVD prepared NFO samples \cite{Meier:2013a}. Since the pure sputter-deposited NFO film (without Pt layer) is insulating within the accuracy of our resistance measurement, we can exclude any ANE coming from the NFO material itself. Therefore, we check the Pt/NFO bilayer for a MPE and, thus, for a MPE induced ANE.

\begin{figure}[!t]
\centering
\includegraphics[width=3in]{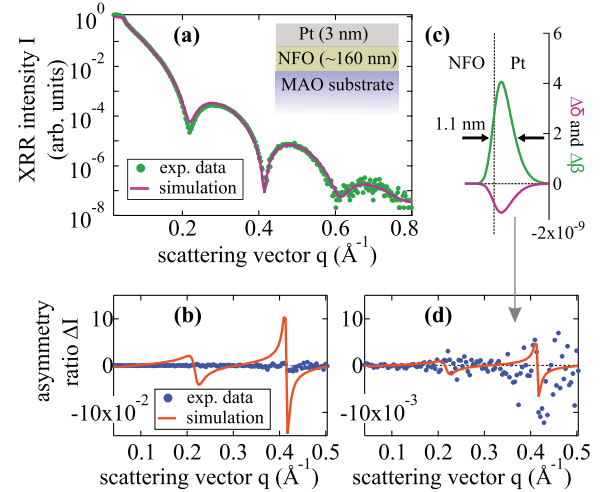}
\caption{(a)~Resonant non-magnetic XRR ($11565\,\textrm{eV}$) intensity $I$ and fit for Pt/NFO. (b)~Corresponding asymmetry ratio $\Delta I$ (average of 8 curves). The same magnetooptic profile as for Pt/Fe($6\,\textrm{nm}$) from Fig.~\ref{fig:Figure2}(h) was used for the simulation. (d)~Close-up of the $\Delta I(q)$ and simulation assuming 5\% of the Pt/Fe($6\,\textrm{nm}$) spin polarization as sketsched in the magnetooptic profile in (c).}
\label{fig:Figure4}
\end{figure}

The resonant XRR curve of Pt/NFO in Fig.~\ref{fig:Figure4}(a) does only show oscillations of the Pt layer due to the large thickness of the NFO of about $(160\pm10)\,\textrm{nm}$ (determined using the calibrated deposition rate). Therefore, the NFO acts quasi as a substrate and the NFO/MAO interface is not accessible as discussed before \cite{Kuschel:2015}. We still can determine a $3\,\textrm{nm}$ thick Pt film with a roughness of $0.4\,\textrm{nm}$ and a Pt/NFO interface with a roughness of $0.3\,\textrm{nm}$. The whiteline intensity of the absorption edge is 1.35 which is slightly above the whiteline intensities obtained for the Pt/Fe samples. The measured XRMR asymmetry ratio in Fig.~\ref{fig:Figure4}(b) is compared to a simulation for the Pt/NFO sample with the same magnetooptic profile of the spin polarization of the Pt/Fe($6\,\textrm{nm}$) interface. The simulated asymmetry ratio is different compared to Pt/Fe, since the optical constants of Fe and NFO vary. However, this curve clearly cannot be identified in the measured data. The larger noise at $q=0.2\,\textrm{\AA}^{-1}$ and $q=0.4\,\textrm{\AA}^{-1}$ is due to the reduced intensity in the XRR curve at these positions. Compared to an asymmetry ratio using a magnetooptic profile with $5\%$ of the Pt/Fe($6\,\textrm{nm}$) spin polarization (cf. Fig.~\ref{fig:Figure4}(c),(d)) we can identify a lower detection limit which leads to a maximum magnetic moment of $0.04\,\mu_{B}$ taking into account the degree of circular polarization of the x-rays and the 70\% magnetized state of the NFO for $\pm200\,\textrm{mT}$. Any MPE in this Pt/NFO bilayer can be neglected down to that limit.

\section{Conclusion}
In conclusion, we used XRMR to investigate the interface spin polarization in Pt/Fe($x$) and Pt/NFO bilayers. We can confirm that the Pt spin polarization is only induced by the interface properties of the magnetic material, since the effect is independent from the thickness of the magnetic layer. We further observed no MPE in Pt on sputter-deposited NFO down to a limit of $0.04\,\mu_{B}$ per Pt atom and, therefore, excluded the MPE induced ANE down to that limit for our LSSE measurements in this bilayer system. These results are comparable to previously investigated Pt/NFO bilayers with chemically prepared NFO \cite{Kuschel:2015}.

\section*{Acknowledgment}
The authors acknowledge financial support by the DFG within the priority program SpinCaT (SPP 1538) via KU~3271/1-1 and RE~1052/24-2.


\begin{thebibliography}{30}

\bibitem{Uchida:2010a}
{K.~Uchida, H.~Adachi, T.~Ota, H.~Nakayama, S.~Maekawa, E.~Saitoh, Appl. Phys. Lett.~\textbf{97}, 172505 (2010).}

\bibitem{Hoffmann:2015}
{A.~Hoffmann, S.~D.~Bader, Phys. Rev. Applied~\textbf{4}, 047001 (2015).}

\bibitem{Bauer:2012}
{G.~E.~W.~Bauer, E.~Saitoh, B.~J.~van~Wees, Nature Mater.~\textbf{11}, 391 (2012).}

\bibitem{Huang:2011}
{S.~Y.~Huang, W.~G.~Wang, S.~F.~Lee, J.~Kwo, C.-L.~Chien, Phys. Rev. Lett.~\textbf{107}, 216604 (2011).}

\bibitem{Avery:2012}
{A.~D.~Avery, M.~R.~Pufall, B.~L.~Zink, Phys. Rev. Lett.~\textbf{109}, 196602, (2012).}

\bibitem{Schmid:2013}
{M.~Schmid, S.~Srichandan, D.~Meier, T.~Kuschel, J.-M.~Schmalhorst, M.~Vogel, G.~Reiss,  C.~Strunk, C.~H.~Back, Phys. Rev. Lett.~\textbf{111}, 187201 (2013).}

\bibitem{Meier:2013b}
{D.~Meier, D.~Reinhardt, M.~Schmid, C.~H.~Back, J.-M.~Schmalhorst, T.~Kuschel, G.~Reiss, Phys. Rev. B~\textbf{88}, 184425 (2013).}

\bibitem{Shestakov:2015}
{A.~S.~Shestakov, M.~Schmid, D.~Meier, T.~Kuschel, C.~H.~Back, Phys. Rev. B~\textbf{92}, 224425 (2015).}

\bibitem{Soldatov:2014}
{I.~V.~Soldatov, N.~Panarina, C.~Hess, L.~Schultz, R.~Sch\"afer, Phys. Rev. B \textbf{90}, 104423 (2014).}

\bibitem{Meier:2015}
{D.~Meier, D.~Reinhardt, M.~van~Straaten, C.~Klewe, M.~Althammer, M.~Schreier, S.~T.~B.~Goennenwein, A.~Gupta, M.~Schmid, C.~H.~Back, J.-M.~Schmalhorst, T.~Kuschel, G.~Reiss, Nat. Commun. \textbf{6}, 8211 (2015).}

\bibitem{Meier:2013a}
{D.~Meier, T.~Kuschel, L.~Shen, A.~Gupta, T.~Kikkawa, K.~Uchida, E.~Saitoh, J.-M.~Schmalhorst, G.~Reiss, Phys. Rev. B~\textbf{87}, 054421 (2013).}

\bibitem{Ramos:2013}
{R.~Ramos, T.~Kikkawa, K.~Uchida, H.~Adachi, I.~Lucas, M.~H.~Aguirre, P.~Algarabel, L.~Morellón, S.~Maekawa, E.~Saitoh, M.~R.~Ibarra, Appl. Phys. Lett. \textbf{102}, 072413 (2013).}

\bibitem{Vlietstra:2014}
{N.~Vlietstra, J.~Shan, B.~J.~van Wees, M.~Isasa, F.~Casanova, J.~Ben~Youssef, Phys. Rev. B \textbf{90}, 174436 (2014).}

\bibitem{Siegel:2014}
{G.~Siegel, M.~C.~Prestgard, S.~Teng, A.~Tiwari, Sci. Rep \textbf{4}, 4429 (2014).}

\bibitem{Gepraegs:2014}
{S.~Gepr\"ags, A.~Kehlberger, F.~Della~Coletta, Z.~Qiu, E.-J.~Guo, T.~Schulz, C.~Mix, S.~Meyer, A.~Kamra, M.~Althammer, H.~Huebl, G.~Jakob, Y.~Ohnuma,	H.~Adachi,	J.~Barker,	S.~Maekawa, G.~E.~W.~Bauer,	E.~Saitoh,	R.~Gross, S.~T.~B.~Goennenwein, M.~Kl\"aui, Nature Commun. \textbf{7}, 10452 (2016).}

\bibitem{Uchida:2013}
{K.~Uchida, T.~Nonaka, T.~Kikkawa, Y.~Kajiwara, E.~Saitoh, Phys. Rev. B.~\textbf{87}, 104412 (2013).}

\bibitem{Uchida:2010b}
{K.~Uchida, T.~Nonaka, T.~Ota, , E.~Saitoh, Appl. Phys. Lett.~\textbf{97}, 262504 (2010).}

\bibitem{Li:2014}
{P.~Li, D.~Ellsworth, H.~Chang, P.~Janantha, D.~Richardson, F.~Shah, P.~Phillips, T.~Vijayasarathy, M.~Wu, Appl. Phys. Lett. \textbf{105}, 242412 (2014).}

\bibitem{Niizeki:2015}
{T.~Niizeki, T.~Kikkawa, K.~Uchida, M.~Oka, K.~Z.~Suzuki, H.~Yanagihara, E.~Kita, E.~Saitoh, AIP Advances \textbf{5}, 053603 (2015).}

\bibitem{Guo:2015}
{E.~Guo, A.~Herklotz, A.~Kehlberger, J.~Cramer, G.~Jakob, M.~Kl\"aui, Appl. Phys. Lett. \textbf{108}, 022403 (2016).}

\bibitem{Saitoh:2006}
{E.~Saitoh, M.~Ueda, H.~Miyajima, G.~Tatara, Appl. Phys. Lett.~\textbf{88}, 182509 (2006).}

\bibitem{Huang:2012}
{S.~Y.~Huang, X.~Fan, D.~Qu, Y.~P.~Chen, W.~G.~Wang, J.~Wu, T.~Y.~Chen, J.~Q.~Xiao, C.~L.~Chien, Phys. Rev. Lett.~\textbf{109}, 147207 (2012).}

\bibitem{Gepraegs:2012}
{S.~Gepr\"ags, S.~Meyer, S.~Altmannshofer, M.~Opel, F.~Wilhelm, A.~Rogalev, R.~Gross, S.~T.~B.~Goennenwein, Appl. Phys. Lett.~\textbf{101}, 262407 (2012).}

\bibitem{Lu:2013}
{Y.~M.~Lu, Y.~Choi, C.~M.~Ortega, X.~M.~Cheng, J.~W.~Cai, S.~Y.~Huang, L.~Sun, C.~L.~Chien, Phys. Rev. Lett.~\textbf{110}, 147207 (2013).}

\bibitem{Valvidares:2015}
{M.~Valvidares, N.~Dix, M.~Isasa, K.~Ollefs, F.~Wilhelm, A.~Rogalev, F.~Sánchez, E.~Pellegrin, A.~Bedoya-Pinto, P.~Gargiani, L.~E.~Hueso, F.~Casanova, J.~Fontcuberta, arXiv:1510.01080.}

\bibitem{Kuschel:2015}
{T.~Kuschel, C.~Klewe, J.-M.~Schmalhorst, F.~Bertram, O.~Kuschel, T.~Schemme, J.~Wollschl\"ager, S.~Francoual, J.~Strempfer, A.~Gupta, M.~Meinert, G.~G\"otz, D.~Meier, G.~Reiss, Phys. Rev. Lett. \textbf{115}, 097401 (2015).}

\bibitem{Klewe:2014}
{C.~Klewe, M.~Meinert, A.~Boehnke, K.~Kuepper, E.~Arenholz, A.~Gupta, J.-M.~Schmalhorst, T.~Kuschel, G. Reiss, J. Appl. Phys. \textbf{115}, 123903 (2014).}

\bibitem{Brown:2001}
{S.~D.~Brown, L.~Bouchenoire, D.~Bowyer, J.~Kervin, D.~Laundy, M.~J.~Longfield, D.~Mannix, D.~F.~Paul, A.~Stunault, P.~Thompson, M.~J.~Cooper, C.~A.~Lucas, and W.~G.~Stirling, J. Synchrotron Radiat. \textbf{8}, 1172 (2001).}

\bibitem{Bouchenoire:2003}
{L.~Bouchenoire, S.~D.~Brown, P.~Thompson, J.~A.~Duffy, J.~W.~Taylor, M.~J.~Cooper, J. Synchrotron Radiat. \textbf{10}, 172, (2003).}

\bibitem{Macke:2014}
{www.remagx.org, S.~Macke, E.~Goering, J. Phys.: Condens. Matter~\textbf{26}, 363201 (2014).}

\bibitem{Klewe:2015}
{C.~Klewe, T.~Kuschel, J.-M.~Schmalhorst, F.~Bertram, O.~Kuschel, J.~Wollschl\"ager, J.~Strempfer, M.~Meinert, G.~Reiss, arXiv:1508.00379.}

\end{thebibliography}
\end{document}